\documentclass[%
reprint,
groupedaddress,
 amsmath,amssymb,
 aps,
]{revtex4-2}

\usepackage{xcolor}
\usepackage{bm}
\usepackage[mathscr]{eucal}
\usepackage{hyperref}
\usepackage{mathtools}

\usepackage{cleveref}
\crefname{equation}{}{}
\crefname{table}{}{}
\crefname{figure}{}{}

\hypersetup{breaklinks=true}

\begin{document}

\preprint{APS/123-QED}

\title{Field space geometry and nonlinear supersymmetry}

\author{Yu-Tse~Lee}%
\affiliation{%
 Department of Physics, University of California, Santa Barbara, California 93106, USA
}%

\begin{abstract}
We propose a geometric formulation of effective field theories via nonlinear supersymmetry. Nonsupersymmetric particles are embedded in constrained superfields governed by a nonlinear $\sigma$ model, and operators are collected into potentials on the target space. The use of chiral superfields standardizes the treatment of flavor across scalars and fermions, and the minimal jet bundle extension makes invariance under derivative field redefinitions manifest.
\end{abstract}

\maketitle

\section{Introduction}

Physics is independent of parametrization. In effective field theory, this principle arises as the invariance of observables under redefinitions between fields and their spacetime derivatives \cite{Chisholm:1961tha, Kamefuchi:1961sb, Arzt:1993gz}. A handle on this redundancy can be obtained by interpreting fields of different flavors as coordinates on a manifold \cite{Coleman:1969sm, Meetz:1969as, Honerkamp:1971xtx, Ecker:1971xko, Honerkamp:1971sh, Alvarez-Gaume:1981exa, Alvarez-Gaume:1981exv, Boulware:1981ns, Howe:1986vm}, so that physical invariants are unambiguously rendered as tensors thereon \cite{Alonso:2015fsp, Alonso:2016oah}. The geometry of field space, applicable to nonderivative field redefinitions, lends important insight into symmetries \cite{Alonso:2015fsp, Alonso:2016btr, Alonso:2016oah, Cohen:2020xca, Alonso:2021rac}, scattering amplitudes \cite{Nagai:2019tgi, Cohen:2021ucp, Helset:2022tlf}, soft theorems \cite{Cheung:2021yog, Derda:2024jvo}, and renormalization \cite{Alonso:2017tdy, Alonso:2022ffe, Helset:2022pde, Jenkins:2023rtg, Jenkins:2023bls, Gattus:2024ird}. However, even after generalizations to the tangent and jet bundles \cite{Craig:2023wni, Craig:2023hhp, Alminawi:2023qtf}, it is yet unclear how a geometric description of derivative field redefinitions can be formulated, without introducing immense additional structure in the form of spacetime or momentum dependence \cite{Cheung:2022vnd, Cohen:2022uuw, Cohen:2023ekv}.

In the field space formalism, Wilson coefficients organize themselves as invariant tensors upon a derivative expansion. For example, in the nonlinear $\sigma$ model
\begin{equation}
\label{eq:scalar_eft}
    \mathcal{L} \supset g_{a\bar{b}}(\phi, \bar{\phi}) \, \partial^\mu \phi^a \partial_\mu \bar{\phi}^{\bar{b}} - V(\phi, \bar{\phi}) \, ,
\end{equation}
two- and zero-derivative operators constitute a metric $g_{a\bar{b}}$ and a scalar $V$ on the target space respectively \footnote{The scalar fields are chosen to be complex for the sake of exposition.}. The derivative counting breaks down under derivative redefinitions which shuffle operators between tensors, modify the kinetic term $g$, and generate higher-order corrections. Nevertheless, the derivative-independent potential $V$ transforms within itself, an important observation that hints at a systematic organization, if one can eliminate the presence of derivatives in the Lagrangian.

On a separate front, field space geometry has been enlarged to incorporate fermions \cite{Finn:2020nvn, Gattus:2023gep, Assi:2023zid}, with complications due to spin and anticommutativity. Allowing for dependence on scalars, the kinetic terms $k_{p\bar{q}}$ and $\omega_{p\bar{q}a}$ for fermions in
\begin{align}
    \mathcal{L} &\supset - \frac{i}{2} k_{p\bar{q}}(\phi, \bar{\phi}) \, \psi^p \sigma^\mu \overset{\text{\tiny$\leftrightarrow$}}{\partial}_\mu \bar{\psi}^{\bar{q}} \label{eq:fermion_eft} \\
    &\quad + \left [ \frac{i}{2} \omega_{p\bar{q}a}(\phi, \bar{\phi}) \, \psi^p \sigma^\mu \bar{\psi}^{\bar{q}} \, \partial_\mu \phi^a + \text{H.c.} \right ] \, , \notag
\end{align}
are typically first order in derivatives \footnote{We write $\psi^p \sigma^\mu \overset{\text{\tiny$\leftrightarrow$}}{\partial_\mu} \bar{\psi}^{\bar{q}} = \psi^p \sigma^\mu (\partial_\mu \bar{\psi}^{\bar{q}}) - (\partial_\mu \psi^p) \sigma^\mu \bar{\psi}^{\bar{q}}$. We consider $\omega_{p\bar{q}a}$ as a kinetic term because it is related to $k_{p\bar{q}}$ via integration by parts.}. In geometric terms, fermionic field space constitutes the full phase space and possesses a symplectic structure, in contrast to scalars whose field space is half of phase space and Riemannian. Combining the two leaves the physical interpretation of the resulting geometry ambiguous, as exemplified by the freedom to modify the metric by various combinations of four-fermion operators \cite{Derda:2024jvo, Note3}\footnotetext{It is also legal in this geometry to treat four-fermion operators as Grassmann corrections to the fermion mass matrix instead.}. This limitation can be alleviated by a formalism that unifies the treatment of flavor and delegates additional particle properties to other structures.

In this paper, we leverage supersymmetry as a resolution to the above considerations, by embedding ordinary fields in chiral superfields $\Phi^I$ on four-dimensional $\mathcal{N} = 1$ superspace. Different particles are now represented by the same type of coordinate, with fermionic properties accounted for by attaching Grassmann spinors $\theta^\alpha$ and $\bar{\theta}^{\dot{\alpha}}$ to spacetime instead. Moreover, the chiral constraint circumvents the need for derivative-dependent kinetic terms, so that the nonderivative supersymmetric nonlinear $\sigma$ model
\begin{equation}
\label{eq:susy_nlsm}
    \mathcal{L} = \int d^4 \theta \, K(\Phi, \bar{\Phi}) + \left [ \int d^2 \theta \, W(\Phi) + \text{H.c.} \right ] \, ,
\end{equation}
is determined by two scalars, the Kähler potential $K$ and superpotential $W$, whose transformations under derivative superfield redefinitions are better behaved.

To make contact with the original theory, supersymmetry must be broken. We treat an effective field theory at energy scale $E$ as the low-energy manifestation of a supersymmetric extension broken at scale $\sqrt{f} \gg E$ \cite{Volkov:1973ix, Ivanov:1978mx}. The superpartners are realized as composite fields containing the Goldstino and suppressed by $f$, ensuring that they remain hidden where the effective theory is valid. The formalism of constrained superfields \cite{Rocek:1978nb, Lindstrom:1979kq, Casalbuoni:1988xh, Komargodski:2009rz, Kuzenko:2010ef}, in which additional nilpotent conditions are imposed on $\Phi^I$, provides a consistent route to nonlinear supersymmetry that retains a superspace description.

In the following, we delineate the range of effective theories that permit a promotion to the supersymmetric nonlinear $\sigma$ model, and find that all operators the latter would yield linearly can be accommodated. They encompass most terms included in existing fermionic field space frameworks and are summarized in Table \cref{tab:susy_embedding} together with their supersymmetric embeddings. Next, we develop a notion of invariance on the associated superfield space, after coordinates for field derivatives have been introduced to form the jet bundle, as illustrated in Fig. \cref{fig:super_jet_bundle}. Physical invariants of the original theory can then be pinned down as invariant tensors after identifying derivative field redefinitions with their superfield counterparts. Finally, we discuss generalizations and directions for future work.

\begin{table}[tbp]
\renewcommand{\arraystretch}{1.5}
\setlength{\arrayrulewidth}{.2mm}
\setlength{\tabcolsep}{1em}
\centering
\begin{tabular}{c|c}
	\hline\hline
    Operators (and H.c.) & Parameters \\
    \hline\hline
    $V(\phi, \bar{\phi})$ & $f_P \in W$ \\
    \hline
    $\partial^\mu \phi^a \partial_\mu \bar{\phi}^{\bar{b}}$ & $g \in K$ \\
    \hline
    $\psi^p \psi^r$ & $m_{pr} + \mathcal{O}(d_{pr\bar{Q}}) \in W$ \\
    \hline
    $\psi^p \psi^r \bar{\psi}^{\bar{q}} \bar{\psi}^{\bar{s}}$ & $r_{p\bar{q}r\bar{s}} + \mathcal{O}(d_{pr\bar{Q}}) \in K$ \\
    \hline
    $\psi^p \sigma^\mu \partial_\mu \bar{\psi}^{\bar{q}} \, , \, \psi^p \sigma^\mu \bar{\psi}^{\bar{q}} \, \partial_\mu \phi^a$ & $k_{p\bar{q}} \in K$ \\
    \hline\hline
\end{tabular}
\caption{On the left are operators in the effective theory, with derivatives or fermion fields explicitly written and scalar fields otherwise. On the right are parameters in the Kähler potential and superpotential of the supersymmetric extension, defined by an expansion \cref{eq:K_and_W} in the Goldstino and fermion superfields. The left can be promoted to the right via \cref{eq:L_EFT}.}
\label{tab:susy_embedding}
\end{table}

\section{Nonlinear Supersymmetry}

To understand when and how an effective field theory (EFT) affords a derivative-independent superfield description, we work backward and derive the nonlinear form of the supersymmetric theory. Consider a low-energy particle content comprising an arbitrary number of Weyl fermions $\psi^p$ and complex scalars $\phi^a$, with no requirement on whether they are superpartners. Since supersymmetry is broken, there must also be a Goldstino $G$, which we will ensure is hidden. For simplicity, we introduce an $F$-term supersymmetry breaking sector and embed $G$ in a chiral superfield
\begin{equation}
    X(x, \theta) = \frac{G(y)^2}{2F(y)} + \sqrt{2} \theta \, G(y) + \theta^2 \, F(y) \, ,
\end{equation}
where $y^\mu = x^\mu + i \theta \sigma^\mu \bar{\theta}$. This superfield satisfies the constraint $X^2 = 0$. Meanwhile, the fermions and scalars are embedded in chiral superfields
\begin{align}
    Y^p &= \left [ \frac{\psi^p G}{F} - \frac{G^2}{2F^2} F^p \right ] + \sqrt{2} \theta \, \psi^p + \theta^2 \, F^p \, , \\
    Z^a &= \phi^a + \sqrt{2} \theta \left [ i \sigma^\mu \frac{\bar{G}}{\bar{F}} \, \partial_\mu \phi^a \right ] \\
    &\quad + \theta^2 \left [ \frac{\bar{G}^2}{2\bar{F}^2} \, \partial^2 \phi^a - \partial_\nu \left ( \frac{\bar{G}}{\bar{F}} \right ) \bar{\sigma}^\mu \sigma^\nu \left ( \frac{\bar{G}}{\bar{F}} \right ) \partial_\mu \phi^a \right ] \, , \notag
\end{align}
that obey $XY^p = 0$ and $X \bar{D}_{\dot{\alpha}} \bar{Z}^a = 0$. Together, these superfields chart a complex manifold $\mathcal{M} = \{ (\Phi^I) \}$, with $I = 0, p, a$ for $X, Y^p, Z^a$ respectively. We collectively denote the component fields by $\phi^I, \psi^I, F^I$ and use the combined index $P \in \{ 0 \} \cup \{ p \}$ for the Goldstino and fermions. The point $\Phi^I = 0$ on $\mathcal{M}$ will be called the vacuum.

Starting from an otherwise linearly supersymmetric theory \cref{eq:susy_nlsm} in $\Phi^I$, the enforcement of constraints on $X$, $Y^p$, and $Z^a$ below $f$ removes the respective superpartners and auxiliary fields $F^a$. They are replaced by composite Goldstino fields as above, so that our desired low-energy degrees of freedom remain, together with $G$ and auxiliary fields $F^P$. Note that the details of supersymmetry breaking are unimportant for our purposes, as long as the same low-energy theory is reproduced. For instance, the constraint $X \bar{D}_{\dot{\alpha}} \bar{Z}^a = 0$ is equivalent to
\begin{equation}
    |X|^2 \bar{D}_{\dot{\alpha}} \bar{Z}^{\bar{a}} = 0 \enspace \text{and} \enspace |X|^2 \bar{D}^2 \bar{Z}^{\bar{a}} = 0 \, ,
\end{equation}
which eliminate the fermion and auxiliary field from $Z^a$ respectively \cite{DallAgata:2016syy}. We can elect not to implement the second constraint \cite{DallAgata:2015zxp}, but any difference, up to suppression by $f$, can be negated by adjusting specific parameters within the superfield Lagrangian.

The nilpotent constraints restrict the forms of the Kähler potential and superpotential below $f$. Notably, cubic terms in $Y^p$ vanish because
\begin{equation}
    X^2 = 0 \enspace \text{and} \enspace X Y^p = 0 \implies Y^p Y^r Y^s = 0 \, .
\end{equation}
Their most general expansions thus read
\begin{subequations}
\begin{align}
    K &= g + b_0 \, \bar{X} X + k_{p\bar{q}} \, \bar{Y}^{\bar{q}} Y^p + r_{p\bar{q}r\bar{s}} \, \bar{Y}^{\bar{q}} \bar{Y}^{\bar{s}} Y^p Y^r \label{eq:Kahler_potential} \\
    &\quad + \Big [ a_0 \, X + a_p \, Y^p + b_p \, \bar{X} Y^p + c_{pr} \, Y^p Y^r \notag \\
    &\qquad + d_{pr\bar{0}} \, \bar{X} Y^p Y^r + d_{pr\bar{q}} \, \bar{Y}^{\bar{q}} Y^p Y^r + \text{H.c.} \Big ] \, , \notag \\
    W &= w + f_0 \, X + f
    _p \, Y^p + m_{pr} \, Y^p Y^r \, , \label{eq:superpotential}
\end{align}
\label{eq:K_and_W}%
\end{subequations}
where lowercase functions only depend on $Z^a$ or $\bar{Z}^{\bar{b}}$.

Taking $f_0(0)$ to be real without loss of generality, we require for consistency that
\begin{equation}
    \frac{f_0(0)}{b_0(0, \bar{0})} \eqqcolon f \, ,
\end{equation}
be large. Then the auxiliary fields $F^P$ can be integrated out straightforwardly by imposing their equations of motion
\begin{equation}
    \kappa_{P\bar{Q}} \, F^P - \frac{1}{2} \kappa_{P\bar{Q},R} \, \psi^P \psi^R + \overline{W}_{,\bar{Q}} = \mathcal{O} \left ( \frac{G}{f} \right ) \enspace \text{and H.c.} \, ,
\end{equation}
where commas indicate partial derivatives on $\mathcal{M}$ and $\kappa_{P\bar{Q}}$ is the submatrix $K_{,P\bar{Q}}$ of the Kähler metric. Goldstino couplings that arise from the removed component fields and possibly contain derivatives become subleading, so that $F^P$ is strongly classical with $\langle F \rangle \sim - f$ in particular. Let us write $k^{\bar{q}p}$ and
\begin{equation}
    \kappa^{\bar{Q}P} = \frac{1}{b'_0} \begin{pmatrix}
        1 & - \bar{b}_{\bar{s}} k^{\bar{s}p} \\
        - k^{\bar{q}r} b_r & b'_0 k^{\bar{q}p} + k^{\bar{q}r} b_r \bar{b}_{\bar{s}} k^{\bar{s}p} \\
    \end{pmatrix} + \mathcal{O} \left ( \frac{G}{f} \right ) \, ,
\end{equation}
for the inverses of $k_{p\bar{q}}$ and $\kappa_{P\bar{Q}}$ respectively, where $b'_0 = b_0 - \bar{b}_{\bar{s}} k^{\bar{s}r} b_r$. We arrive at a nonlinearly supersymmetric theory
\begin{equation}
    \mathcal{L} = \mathcal{L}_{\text{EFT}} + \mathcal{L}_G \, ,
\end{equation}
in terms of the low-energy particles and the Goldstino only, where
\begin{widetext}
\begin{align}
    \mathcal{L}_{\text{EFT}} &= g_{,a\bar{b}} \, \partial^\mu \phi^a \partial_\mu \bar{\phi}^{\bar{b}} - \kappa^{\bar{Q}P} f_P \bar{f}_{\bar{Q}} - \frac{i}{2} k_{p\bar{q}} \, \psi^p \sigma^\mu \overset{\text{\tiny$\leftrightarrow$}}{\partial}_\mu \bar{\psi}^{\bar{q}} \label{eq:L_EFT} \\
    &\quad + \left [ \frac{i}{2} k_{p\bar{q},a} \, \psi^p \sigma^\mu \bar{\psi}^{\bar{q}} \, \partial_\mu \phi^a - \left ( m_{pr} - \kappa^{\bar{Q}P} d_{pr\bar{Q}} \, f_P \right ) \psi^p \psi^r + \text{H.c.} \right ] + \left [ r_{p\bar{q}r\bar{s}} - \kappa^{\bar{Q}P} d_{pr\bar{Q}} \, \bar{d}_{\bar{q}\bar{s}P} \right ] \psi^p \psi^r \bar{\psi}^{\bar{q}} \bar{\psi}^{\bar{s}} \, , \notag \\
    \mathcal{L}_G &= - \frac{i}{2} b_0 \, G \sigma^\mu \overset{\text{\tiny$\leftrightarrow$}}{\partial}_\mu \bar{G}^{\bar{q}} - \frac{i}{2} b_p \, \psi^p \sigma^\mu \overset{\text{\tiny$\leftrightarrow$}}{\partial}_\mu \bar{G} - \frac{i}{2} \bar{b}_{\bar{q}} \, G \sigma^\mu \overset{\text{\tiny$\leftrightarrow$}}{\partial}_\mu \bar{\psi}^{\bar{q}} \label{eq:L_G} \\
    &\quad + \left [ \frac{i}{2} b_{0,a} \, G \sigma^\mu \bar{G} \, \partial_\mu \phi^a + \frac{i}{2} b_{p,a} \, \psi^p \sigma^\mu \bar{G} \, \partial_\mu \phi^a + \frac{i}{2} \bar{b}_{\bar{q},a} \, G \sigma^\mu \bar{\psi}^{\bar{q}} \, \partial_\mu \phi^a + \text{H.c.} \right ] + \mathcal{O} \left ( \frac{G}{f} \right ) \, . \notag
\end{align}
\end{widetext}
The Goldstino couplings include unsuppressed kinetic terms determined by $b_P$, which we therefore demand be small.

With all Goldstino couplings now accounted for, we conclude that any low-energy theory of the form \cref{eq:L_EFT} has a derivative-independent supersymmetric extension \cref{eq:susy_nlsm} subject to nilpotent constraints. The extension is obtained by inserting operators into the potentials according to \cref{eq:K_and_W}, and appending a hidden Goldstino sector \cref{eq:L_G}. Sensibly, all operators of the elementary fields that would appear in a linearly supersymmetric nonlinear $\sigma$ model assume their usual places, and new operators involving the composite superpartners are suppressed.

The cost of uplifting to a superfield description is relatively modest. The starting visible theory need not be linearly supersymmetric, and the parameters $f$ and $b_P$ can be tuned to control the size of any additional operator introduced. A comparison between \cref{eq:scalar_eft} and \cref{eq:L_EFT} suggests the possibility of global obstructions due to the scalar kinetic term and potential. However, near the vacuum on $\mathcal{M}$, a local Kähler potential and holomorphic square root generally exist for the two quantities respectively. In the context of effective field theory, the primary restriction, apart from the allowable classes of operators, is that the fermion kinetic terms in \cref{eq:fermion_eft} must satisfy $\omega_{p\bar{q}a} = k_{p\bar{q},a}$ \footnote{That is, the vector bundle of fermions over scalars has a Chern connection.}.

In exchange, we are compensated by several practical gains. Rather than keeping track of various theory parameters on ordinary field space, we can now organize them into two superfield potentials. It is fairly straightforward to read the parameters off from a superfield expansion, given the simple correspondence between the low-energy fields and their superfield embeddings. Moreover, there is now an unequivocal identification of four-fermion operators with the visible piece of the curvature on $\mathcal{M}$, given by the last term in \cref{eq:L_EFT}.

Conceptually, the superfield space $\mathcal{M}$ provides an egalitarian treatment of particle flavor regardless of spin and anticommutativity. Such properties are instead carried by the superspace $\mathcal{T}$ \footnote{Nilpotent constraints on superfields operate on component fields and invoke the structure of the supermanifold $\mathcal{T}$, not the regular manifold $\mathcal{M}$.}. Arguably, the geometry of $\mathcal{M}$ is physically significant since all parameters in $K$ and $W$ relevant to the low-energy theory have been fixed \footnote{Kähler transformations do not produce physical effects on the component theory.}. And importantly, the superfield formalism makes invariance under derivative field redefinitions manifest, as we will soon explain.

We remark that the parameters $a_P$, $c_{pr}$, and $w$ live strictly in the hidden sector. They would have been important if one were interested in the Goldstino couplings. We also comment that the framework can be generalized to real scalar theories, in which case an additional constraint
\begin{equation}
    |X|^2 B^a = 0 \enspace \text{where} \enspace Z^a = A^a + i B^a \, ,
\end{equation}
eliminates the imaginary component. Having limited ourselves to complex scalars, we reap the benefit of simplifying the geometric machinery in the next section.

\section{Field Redefinition Invariance}

The supersymmetric extension of an effective field theory we constructed standardizes particles as superfields and packages Wilson coefficients as two potentials on the superfield space $\mathcal{M}$. Besides the organizational advantage, the new description is also derivative independent. Unlike the typical spacetime Lagrangian which is nontrivial in fields and field derivatives, the superspace Lagrangian resides within a special slice. The resulting geometry is more tractable and manifests invariance under derivative component field redefinitions, which we implement via their superfield counterparts.

We begin with nonderivative superfield redefinitions $\Phi^I(\Phi'^J)$ on $\mathcal{M}$ \footnote{The redefinitions are holomorphic, in accordance with the complex geometry of $\mathcal{M}$.}, specifically the subset
\begin{equation}
\label{eq:non-deriv_susy}
    Y^p = S^p_r(Z'^c) \, Y'^r \enspace \text{and} \enspace Z^a = Z^a(Z'^c) \, ,
\end{equation}
where $S^p_r$ is an invertible matrix. They respect the division between the visible and hidden sectors. They also happen to preserve the forms of the nilpotent constraints and the expansions of $K$ and $W$. At the component level, they are equivalent to the nonderivative redefinitions
\begin{equation}
\label{eq:non-deriv_eft}
    \psi^p = S^p_r(\phi'^c) \, \psi'^r + \mathcal{O}(G/f) \enspace \text{and} \enspace \phi^a = \phi^a(\phi'^c) \, ,
\end{equation}
which are the coordinate transformations on field space as formulated in most literature. Tensors on superfield space are therefore tensorial on field space. The mapping between \cref{eq:K_and_W} and \cref{eq:L_EFT} yields a direct translation between the two geometries, allowing us to rewrite Wilson coefficients of the effective theory invariantly on $\mathcal{M}$.

As an example, the tree-level scattering amplitude $\psi^r \bar{\psi}^{\bar{s}} \rightarrow \phi^a \bar{\phi}^{\bar{b}}$ of the low-energy theory reads
\begin{equation}
\label{eq:amplitude}
    \mathcal{A} = R^{r\bar{s}a\bar{b}}(0, \bar{0}) \, (p^r_\mu + p^a_\mu) \, \mathsf{x}(p^r) \sigma^\mu \bar{\mathsf{y}}(p^s) \, ,
\end{equation}
where $p^I_\mu$ are the ingoing particle momenta and $\mathsf{x}, \, \bar{\mathsf{y}}$ are two-component spinor wave functions \cite{Dreiner:2008tw, Note8}\footnotetext{This expression can be easily derived in normal coordinates on superfield space \cite{Cohen:2021ucp}, whose Feynman rules are simple compared to generic parameterizations on $\mathcal{M}$ or more generally $\mathcal{J}$.}. Instead of Wilson coefficients, the numbers that specify the amplitude as a function of kinematics are expressed as tensors on $\mathcal{M}$ evaluated at the vacuum, in this case the curvature $R$. The tensors are manifestly invariant under superfield redefinitions like \cref{eq:non-deriv_susy} and hence the corresponding component redefinitions like \cref{eq:non-deriv_eft}.

Going one step beyond, consider derivative superfield redefinitions such as
\begin{gather}
    Y^p = Y'^p + S^p_{qa}(Z'^c) \, \partial^\mu Y'^q \, \partial_\mu Z'^a \\
    \text{or} \enspace Z^a = Z'^a + S^a_b(Z'^c) \, \partial^\mu \partial_\mu Z'^b \, , \notag
\end{gather}
which enact the component redefinitions
\begin{gather}
    \psi^p = \psi'^p + S^p_{qa}(\phi'^c) \, \partial^\mu \psi'^q \, \partial_\mu \phi'^a + \mathcal{O}(G/f) \label{eq:deriv_redef} \\
    \text{or} \enspace \phi^a = \phi'^a + S^a_b(\phi'^c) \, \partial^\mu \partial_\mu \phi'^b \, . \notag
\end{gather}
We wish to rewrite the numbers in the amplitude as objects invariant under the more general redefinitions. The appropriate setting must be an extended manifold charted not just by $\Phi^I$ but also its derivatives, leading us to the jet bundle $\mathcal{J}$ \cite{ehresmann1953, saunders1989, anderson1992}. Defined as the set of equivalence classes of superfields with the same Taylor series over superspace, $\mathcal{J}$ has coordinates
\begin{equation}
    (x^\mu, \, \theta^\alpha, \, \bar{\theta}^{\dot{\alpha}}, \, \Phi^I, \, \partial_\mu \Phi^I, \, \partial_\alpha \Phi^I, \, \partial_{\dot{\alpha}} \Phi^I, \ldots) \, ,
\end{equation}
where even derivatives $\partial_\mu$ can be of arbitrary order while odd derivatives $\partial_\alpha$ and $\partial_{\dot{\alpha}}$ are finite. The superspace coordinates play no vital role in the following discussion, so we subsequently suppress their reference. For brevity, we denote the collection of all derivative coordinates by $\delta \Phi^{\mathtt{I}} = \partial_\mu \Phi^I, \partial_\alpha \Phi^I, \ldots$ and all coordinates by $\Xi^{\mathscr{I}} = \Phi^I, \delta \Phi^{\mathtt{I}}$.

The jet bundle accommodates derivative superfield redefinitions
\begin{equation}
\label{eq:jet_coord_trans}
    \Phi^I = \Phi^I(\Phi', \, \partial_\mu \Phi', \, \partial_\alpha \Phi', \, \partial_{\dot{\alpha}} \Phi', \, \ldots) \, ,
\end{equation}
as coordinate transformations. The chain rule stipulates that derivative coordinates transform correspondingly as
\begin{equation}
    \partial_\mu \Phi^I = \frac{\partial \Phi^I}{\partial \Phi'^J} \, \partial_\mu \Phi'^J + \frac{\partial \Phi^I}{\partial (\partial_\nu \Phi'^J)} \, \partial_\mu \partial_\nu \Phi'^J + \ldots \, ,
\end{equation}
and so on. Thus, the Jacobian takes the form
\begin{equation}
    M^{\mathscr{I}}_{\mathscr{J}} \coloneqq \frac{\partial (\Phi^I, \, \delta \Phi^{\mathtt{I}})}{\partial (\Phi'^J, \, \delta \Phi'^{\mathtt{J}})} = \begin{pmatrix}
        \mu^I_J & \bigtimes \\
        \mathcal{O}(\delta \Phi) & \bigtimes
    \end{pmatrix} \, .
\end{equation}
The second column is inconsequential to us \footnote{Besides invertibility requirements which are indeed satisfied.}, the submatrix $\partial (\delta \Phi^{\mathtt{I}}) / \partial \Phi'^J$ has support on derivative coordinates \footnote{Note that $\delta \Phi^{\mathtt{I}} = 0 \iff \delta \Phi'^{\mathtt{J}} = 0 \,$.}, and the submatrix $\mu^I_J = \partial \Phi^I / \partial \Phi'^J$ reduces to the superfield space Jacobian when the redefinition is nonderivative.

Denote the submanifold of $\mathcal{J}$ where all derivative coordinates vanish as $\mathcal{E} \cong \mathcal{T} \times \mathcal{M}$. Among the holomorphic tangent vectors on $\mathcal{J}$, those in the superfield directions transform as
\begin{equation}
    \frac{\partial}{\partial \Phi'^J} = M^{\mathscr{I}}_J \frac{\partial}{\partial \Xi^{\mathscr{I}}} = \mu^I_J \frac{\partial}{\partial \Phi^I} + \mathcal{O}(\delta \Phi) \, .
\end{equation}
In particular, they transform among themselves on the restriction of $\mathcal{J}$ to $\mathcal{E}$ and form a subbundle of $T^{1,0}_{\mathcal{E}} \mathcal{J}$. The same holds for the antiholomorphic vectors. We can then use them to build contravariant tensors
\begin{equation}
    T = T^{I \bar{J} K \cdots} \frac{\partial}{\partial \Phi^I} \otimes \frac{\partial}{\partial \bar{\Phi}^{\bar{J}}} \otimes \frac{\partial}{\partial \Phi^K} \otimes \ldots \, ,
\end{equation}
on $\mathcal{J}|_\mathcal{E}$, given components $T^{I \bar{J} K \cdots}$ that transform oppositely.

Let us extend the potentials $K$ and $W$ from $\mathcal{M}$ to $\mathcal{J}$ by taking them to be independent of the derivative coordinates $\delta \Phi^{\mathtt{I}}$ \footnote{As they already are in the superspace Lagrangian.}. This triviality simplifies the transformation laws of their partial derivatives in the superfield directions and of further quantities generated. All transform like they would on $\mathcal{M}$, but with the superfield space Jacobian and its derivatives replaced by the sub-Jacobian $\mu^I_J$ on $\mathcal{J}$ and its derivatives. For instance, the Kähler metric obeys
\begin{equation}
    \frac{\partial K}{\partial \Phi'^K \partial \bar{\Phi}'^{\bar{L}}} = M^{\mathscr{I}}_K \, \overline{M}^{\bar{\mathscr{J}}}_{\bar{L}} \, \frac{\partial K}{\partial \Xi^{\mathscr{I}} \partial \bar{\Xi}^{\bar{\mathscr{J}}}} = \mu^I_K \, \overline{\mu}^{\bar{J}}_{\bar{L}} \, \frac{\partial K}{\partial \Phi^I \partial \bar{\Phi}^{\bar{J}}} \, ,
\end{equation}
and the holomorphic connection coefficients obey
\begin{equation}
    \Gamma'^L_{\,MN} = (\mu^{-1})^L_I \, \mu^J_M \, \mu^K_N \, \Gamma^I_{JK} + (\mu^{-1})^L_I \frac{\partial \Phi^I}{\partial \Phi'^M \partial \Phi'^N} \, .
\end{equation}
Contravariant tensors on $\mathcal{M}$, derived from the two potentials and extended to $\mathcal{J}$ as above, thus provide the components we need.

\begin{figure}[tbp]
\centering
\includegraphics[width=\columnwidth]{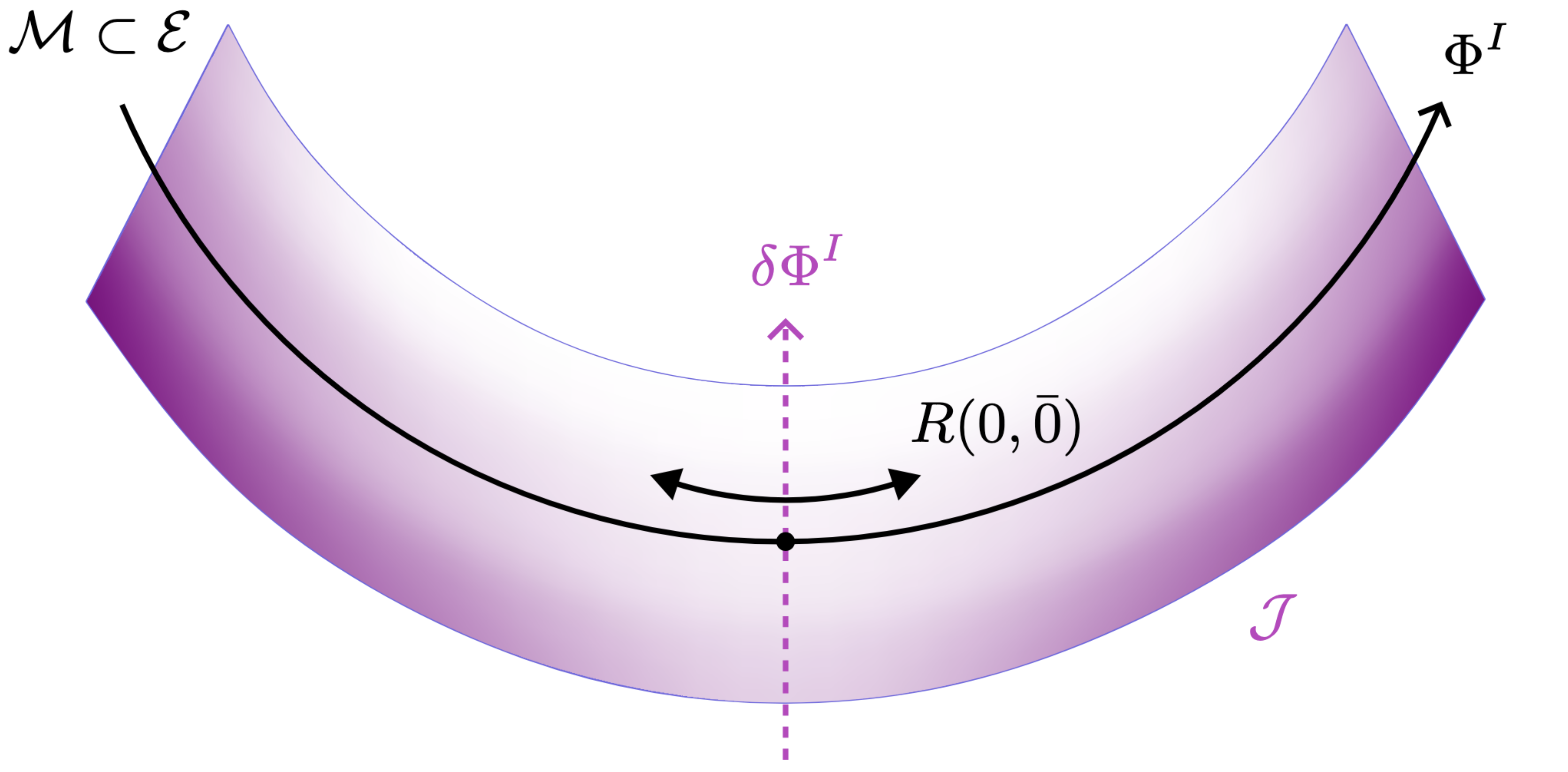}
\caption{\label{fig:super_jet_bundle}The geometry of superfield space $\mathcal{M}$, charted by $\Phi^I$, extends trivially when derivative coordinates $\delta \Phi^I$ are introduced to form the jet bundle $\mathcal{J}$. Geometric quantities on $\mathcal{M}$, such as the curvature in the scattering amplitude \cref{eq:amplitude}, remain invariant under the larger set of superfield redefinitions \cref{eq:jet_coord_trans} now permitted.}
\end{figure}

As depicted in Fig. \cref{fig:super_jet_bundle}, we have extended the geometry of superfield space $\mathcal{M}$ in the obvious way to $\mathcal{E}$ and $\mathcal{J}$, introducing no dependence on superspace or derivative coordinates. Triviality in the latter ensures that the extension is tensorial even under derivative superfield redefinitions. Returning to amplitudes, all it takes to make $\cref{eq:amplitude}$ manifestly invariant under \cref{eq:deriv_redef} is to uplift the tensors on $\mathcal{M}$ to their immediate extensions on $\mathcal{J}|_{\mathcal{E}}$. Essentially, the superfield space geometry derived from $K$ and $W$ is automatically invariant under a larger set of coordinates and so are the amplitudes it renders \footnote{This is unlike the geometry of the ordinary jet bundle derived from a field space connection \cite{Craig:2023hhp}, which is only invariant under non-derivative field redefinitions.}.

To be clear, the geometric machinery applies to all reparametrizations of superfields, which are governed by a derivative-independent Lagrangian. Meanwhile, the supersymmetric embedding in the previous section is needed to make the translation to fields in the original derivative-dependent theory. In particular, redefinitions of the fermions $\psi^p$ are implemented via $Y^p$ and necessarily modify the auxiliary fields $F^p$ as well. This apparent issue can be circumvented since $F^p$ does not generate any new non-Goldstino operator over $F$ in \cref{eq:L_EFT} and hence can be made entirely decoupled from the visible sector.

More generally, the space of derivative superfield redefinitions exceeds that of the starting low-energy fields $\psi^p$ and $\phi^a$, since new component fields have been introduced after all. Redefinitions involving the Goldstino superfield $X$ can spoil the division between the visible and hidden sectors. In addition, redefinitions like
\begin{equation}
    Z^a = Z'^a + S^a_{pq}(Z'^c) \, D^\alpha Y^p D_\alpha Y^q \, ,
\end{equation}
which would have mixed fermions into scalars, make the new superfield no longer chiral. All of them are nevertheless valid redefinitions over larger sets of component fields than the original ones.

Although superfields are needed as a proxy, the present construction offers an advantage in tractability over uncountably sized structures that can handle derivative field redefinitions. As a framework that unifies the treatment of scalar and fermion flavors, it explicitly accounts for derivative redefinitions involving the latter. And by maintaining a close relationship with field space geometry, the invariant quantities it distills from a theory are more amenable to physical interpretation.

\section{Outlook}

The vast reparametrization redundancies in effective field theories are difficult to tease apart, with further complexities upon generalizing to fermions and field derivatives. The application of nonlinear supersymmetry to field space geometry addresses the two problems at once. It yields the same representation for fields of different spins and standardizes their treatment. Moreover, it organizes a range of operators into potentials and simplifies the associated geometry. The framework of superfield space handles field redefinitions spanning multiple particle types and a qualitatively larger set of coordinates.

Regarding field redefinitions, an immediate extension pertains to operators beyond those of the supersymmetric nonlinear $\sigma$ model. Most of them can be accommodated through derivative corrections to the superspace Lagrangian \cite{Antoniadis:2007xc, Khoury:2010gb}, which fit within $\mathcal{M}$ as tensors rather than scalars. The superfield description retains most of its merits, but invariance under derivative redefinitions is no longer obvious. There are other generalizations that motivate further geometric constructions, such as the incorporation of gauge fields \cite{Helset:2018fgq, Helset:2022tlf, Helset:2022pde} using vector multiplets.

As for supersymmetry, its connection to effective field theories extends beyond field redefinitions. As a candidate model of nature, it registers observable effects when broken at a lower energy scale, through Goldstino couplings that can be worked out using a similar approach \cite{Antoniadis:2010hs, Dudas:2012fa}. Apart from possible direct relevance, it also serves as a toolbox to understand more formal structures, such as holomorphicity and nonrenormalization results \cite{Alonso:2014rga, Elias-Miro:2014eia}. There is more to be learned about generic theories through the lens of supersymmetry.

\vspace{1em}

\noindent \textit{Note added:} Recently, the author learned of the concurrent work by Cohen, Lu, and Zhang \cite{Cohen:2024bml}, who have provided a new geometric formula that makes manifest the on-shell covariance of amplitudes under general field redefinitions. The author thanks them for sharing their draft and their kind correspondence.

\vspace{1em}

\section*{Acknowledgments}

The author is grateful to Nathaniel Craig and Andrew Fee for collaboration on related topics and valuable comments on the manuscript, and Maria Derda and Julio Parra-Martinez for helpful discussions. This work was supported in part by the U.S. Department of Energy under the Grant No. DE-SC0011702.

\bibliography{arxiv_v3}

\end{document}